\documentclass{elsart}

\usepackage{amsmath}                   % Use AMS math environments.
\usepackage{amssymb}                   % Use AMS symbols and fonts.
\usepackage{graphicx}                  % Graphics include environment.
\usepackage{lineno}                    % Package to number all lines.
\begin{document}

% \linenumbers                           % Comment out for no line numbering.

% -----------------------------------------------------------------------------
\begin{frontmatter}
\journal{Nucl. Instr. Meth. A}

\title{
   A Prototype PCI-based Data Acquisition \\
   System for Cosmic Ray Detection\\
   Below $10^{18}$~eV
   }

\author{S. BenZvi\corauthref{cor1}},
\corauth[cor1]{Corresponding author.  Tel: +1-212-854-2363; fax: +1-212-854-3379}
\ead{sybenzvi@phys.columbia.edu}
\author{S. Westerhoff},
\author{J. Ban},
\author{W.F. Sippach}
\address{Columbia University, Department of Physics, Nevis Laboratories,
136 S.~Broadway, P.O. Box 137, Irvington, NY 10533, USA}

% -----------------------------------------------------------------------------

\begin{abstract}
A prototype flash analog-to-digital readout system for cosmic ray detection 
at energies below $10^{18}$~eV has been designed and tested at Columbia 
University Nevis Laboratories.  The electronics consist of a FADC module that 
digitizes 16 photomultipliers at 40~MHz with 14-bit dynamic range.  The module 
is read out to a PC (running Linux) through a PCI interface.  Taking advantage 
of the large bandwidth provided by the PCI bus, we have implemented a 
software-based data acquisition system.  This note describes the software and
electronics, as well as preliminary tests carried out using a prototype FADC
module.
\end{abstract}


\end{frontmatter}

% -----------------------------------------------------------------------------
\section{Introduction}
% -----------------------------------------------------------------------------
In the past few years, several proposals have been made to extend the
observation of ultra high-energy cosmic rays (UHECRs) to energies well below 
$10^{18}$~eV~\cite{Adams:2003sk,Haungs:2004gq}, the threshold of currently 
operated UHECR detectors.  Previous measurements have indicated that at these 
energies, the cosmic ray energy spectrum, a steeply falling power law in energy
$E^{-\alpha}$, with spectral index $\alpha\simeq 3$, shows structure, 
and the chemical composition undergoes a change from a heavier, iron-dominated 
mixture to a lighter, proton-dominated composition~\cite{Abu-Zayyad:2000ay}.

Changes in composition and in the index of the energy spectrum could 
indicate a transition in the sources of cosmic rays, from a Galactic origin at 
lower energies to an extragalactic origin at higher energies.  The transition
region, between the ``knee'' of the spectrum --- a steepening of the index
around $10^{15}~\text{eV}$~\cite{Haungs:2004gq} --- and the start of the ultra
high-energy regime above $10^{18}~\text{eV}$, has not been observed in detail.
The basic requirements for a detector measuring cosmic rays at such energies 
are excellent energy resolution, good angular resolution, and the ability to 
discriminate between different cosmic ray primaries, at least on a statistical
basis.  

A measurement technique that meets these requirements is the air fluorescence
method, in which a detector observes the nitrogen fluorescence light generated
by secondary air shower particles in the atmosphere.  Most of the fluorescence
light is emitted at wavelengths between $300~\text{nm}$ and $450~\text{nm}$,
requiring a UV-sensitive instrument.  Typically, an air fluorescence 
detector is composed of a large number of optically fast ``cameras,'' each 
collecting shower light with a wide-angle mirror, and using several hundred
photomultipliers (PMTs) with fast electronics to image the developing shower in
real time.  

The fluorescence signal is weak compared to other light sources,
such as moonlight, but can be observed over a dark sky background.  To maximize
signal to noise, the camera mirrors tend to be large, on the order of a few
square meters, and the PMTs observe a relatively small portion of
the sky (usually $\sim1^{\circ}\times1^{\circ}$).  Currently operating cosmic
ray detectors that employ the air fluorescence technique are the High
Resolution Fly's Eye (HiRes) experiment in Utah~\cite{Boyer:2002zz}, and the 
Pierre Auger Observatory in Argentina~\cite{Dova:2001aw}.

Since air fluorescence detectors observe extensive air showers as they
develop in the atmosphere, the technique yields not only accurate measurements
of shower geometry, but also calorimetric estimates of the primary particle 
energy.  However, the measurements are also plagued by several major technical
difficulties.  The fluorescence signal strength from cosmic ray air showers 
varies greatly due to shower-to-shower fluctuations.  In HiRes, for example,
the typical shower signal ranges between 200 photoelectrons to several 
thousand in a $100~\text{ns}$ time window~\cite{Boyer:2002zz}.  Moreover, the 
signal sits atop a slowly varying dark sky background --- on the order of 
tens of photoelectrons per $\mu\text{s}$ per PMT --- that can increase by an
order of magnitude when a bright star crosses the PMT
border~\cite{Boyer:2002zz}.  Finally, while the duration of each shower is 
short, of order $\mu\text{s}$, event rates are high, on the order of kHz.

The major challenges posed by extending the air fluorescence technique 
below $10^{18}~\text{eV}$ include the large dynamic range of the fluorescence 
and background signals, and the large increase in rate at lower energies.
To accommodate such observations, we have designed and partially implemented a 
fully digital readout system for an air fluorescence telescope.  In its current
form, the readout system, which has been tested at Columbia University Nevis 
Laboratories, contains three basic components: a subcluster of sixteen Photonis
XP3062 photomultipliers; an ``FADC module'' responsible for digitizing the PMT 
outputs and making basic trigger decisions; and a compact PCI board that 
handles two-way communication between the FADC electronics and a data 
acquisition (DAQ) PC running Linux.  

In this paper, we discuss the components of the electronics system in detail, 
and describe the results of basic calibrations.  The paper is organized as 
follows.  Sections~\ref{subsec:design} and \ref{subsec:system} give a brief 
description of the design consideration for the development of a DAQ system for
air fluorescence detectors at lower energies.  We then describe the FADC module
(Section~\ref{subsec:fadcmodule}), the PCI readout board 
(Section~\ref{subsec:pci}), and the data acquisition system 
(Section~\ref{subsec:daq}).  Some benchmarks are discussed and summarized in 
Section~3.

% -----------------------------------------------------------------------------
\section{Description of the System}
% -----------------------------------------------------------------------------

\subsection{Design Considerations}\label{subsec:design}
In the design of the readout electronics, we attempted to follow three guiding
principles.  First, so as to limit analog noise and ease the signal processing
requirements, the electronics are set up to digitize the PMT outputs 
immediately after integration and shaping by two preamplifiers.  All subsequent
monitoring and triggering tasks are performed on the digitized waveforms.  
Second, the hardware controller does not require a cumbersome VME interface;
access occurs through a custom designed PCI card using simple function calls.
Third, the system offloads most trigger decisions to the DAQ software, taking 
advantage of the speed of the DAQ PC.  This greatly simplifies the overall 
requirements of the electronics and guarantees maximum flexibility in the 
implementation of various trigger schemes.  The large bandwidth of the PCI bus
($130$\,MB\,s$^{-1}$ nominal) easily accommodates the large flow of data from 
the FADC module.

% -----------------------------------------------------------------------------
\subsection{System Overview}\label{subsec:system}
% -----------------------------------------------------------------------------
A basic outline of the PMT readout chain is as follows.  The PMT bases contain
a simple board to supply HV and read out the anode currents.
After the anode output is integrated by a preamplifier, a high bandwidth
FireWire cable transfers the signal to a 14-bit FADC, where it is digitized at
$40$\,MHz.

Each FADC is located on a board with sixteen channels in total (i.e., one board
reads out sixteen PMTs.)  Monitoring the sixteen channels is an Altera FPGA,
which tracks the channel baselines with a simple baseline finder, and 
implements a simple threshold and coincidence trigger scheme for the PMT
``mini-cluster'' it controls.  The FPGA firmware has also been programmed as an
event builder, packaging the sixteen tube signals into a simple data format ---
word count, time stamp, and channel outputs -- for later processing.  

% Fig 1
\begin{figure}
\centering
   \includegraphics*[width=0.7\textwidth,angle=0,clip]{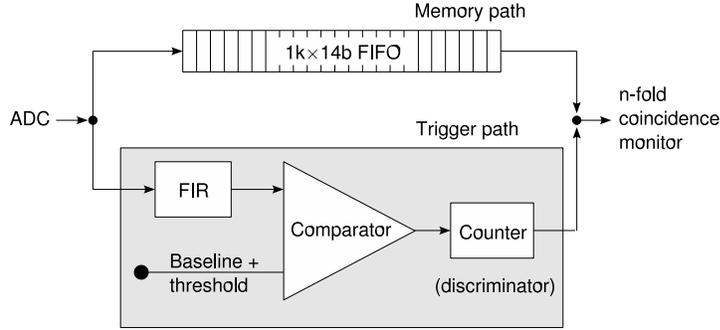}
   \caption{Logical setup of an FADC channel.  In the trigger path, the
      FIR integration time, external baseline, threshold, counter level,
      and n-fold parameter are set by software knobs.}
   \label{fig:adc}
   \vspace{0.5cm}
\end{figure}

In our mini-cluster, the FPGA-FADC board cannot alter the PMT gains.  This 
approach is taken because the large dynamic range of the FADCs allows us to 
avoid using an overflow channel for saturated tubes.  Hence, we do not
normalize the inputs to the board, aside from some course gain adjustments at 
the tube level. 

Once the FPGA packages the FADC signals into a data block, the block moves
over another high-bandwidth connection to a PCI card, which provides the
interface to hard disk storage.  The bandwidth of the PCI bus allows the card
to rapidly and transparently move events from the detector directly to PC
memory, where they are saved in a large buffer.  Once in memory, the data can 
be read from the buffer, whereupon higher-level triggers may be applied to the 
PMT cluster output.

In the following sections, we give a more detailed description of each of the
components of the system.

% -----------------------------------------------------------------------------
\subsection{FADC Module}\label{subsec:fadcmodule}
% -----------------------------------------------------------------------------
The FADC electronics are composed of two PCBs: a single Digital Signal 
Processing (DSP) board that accepts differential analog signals from sixteen 
phototubes, and a backplane that receives i/o and clock input from a PC via the
PCI interface.  Analog data from the sixteen phototubes are transported to the 
DSP board by two high-bandwidth Mini D ribbon cables, while the DSP and 
backplane communicate via three MZP board-to-board PCB plugs.

After arriving at the DSP board, the integrated and shaped analog output from
each tube in the PMT subcluster is processed by two high-speed differential
amplifiers and digitized by a $40$~MHz, $14$~bit, $300$~mW flash 
analog-to-digital converter (FADC).  The particular converter used in the
FADC module (Analog Devices AD9244) was selected for its good balance between
high digitization rate and low power consumption \cite{AD9244:2004b}.

Following digitization by the sixteen FADCs, each input channel splits the
digitized data along two paths, as shown in Fig.~\ref{fig:adc}: a trigger path 
for signal processing by an Altera Stratix FPGA~\cite{Altera:2004}; and a deep 
memory path to store the data while the trigger decision occurs.  The memory 
path is implemented by a $1$k$\times14$b FIFO, which is sufficient to store 
the data for $25~\mu$s during the trigger stage.

% Fig 2
\begin{figure}
\centering
   \includegraphics*[width=0.6\textwidth,angle=0,clip]{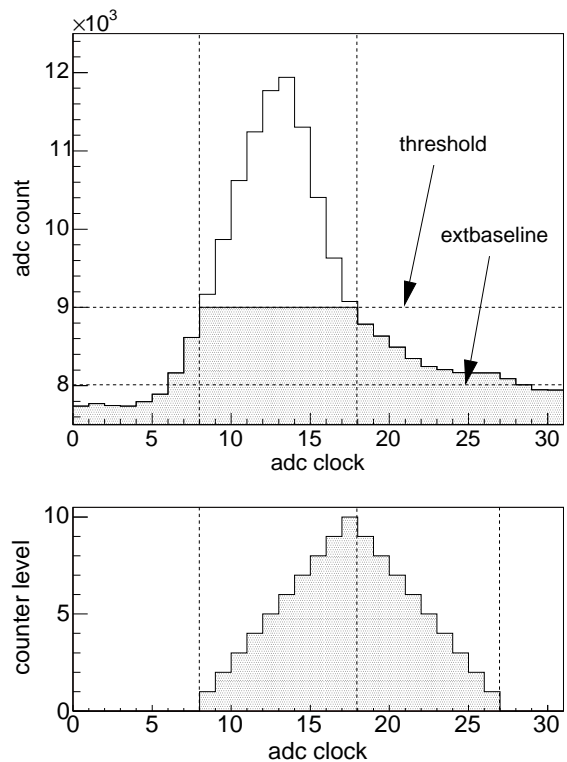}
   \caption{\footnotesize Effect of the channel counter, which doubles the
      length of time the above-threshold signal can be used for coincidence 
      triggers.}
   \label{fig:counter}
   \vspace{0.5cm}
\end{figure}

In the trigger path, the digitized photomultiplier waveforms are integrated
over one, two, four, or eight successive samples by a FIR filter implemented in
the Stratix FPGA.  The integration time of the filter is adjustable by a DAQ 
software knob.  The integrated output is compared to the sum of user-supplied
external baseline and threshold parameters, and then is further discriminated 
by a counter.  The external channel baseline, trigger threshold, and counter 
discriminator window are unique to each channel.  

At each ADC clock, the counter increments when the PMT signal is above 
threshold, and decrements when the signal drops below threshold.  A
discriminator monitors the counter level, and when it exceeds a tunable
base value for a tunable number of clock cycles, the discriminator will enable 
the trigger channel.  Hence, the counter can effectively double the length of 
time available for time coincidence between channels when compared to using 
the signal alone (see Fig.~\ref{fig:counter}).

Note that the counter discriminator, used in this manner, biases air shower
detection toward low-energy events, which tend to occur relatively 
close to the detector and give rise to a strong light signal at the camera, and
against high-energy events, which tend to occur farther away, have a larger 
spread in signal times, and have relatively low light levels.  The bias toward
close showers not only reduces the overall event rate, but also decreases the 
uncertainty caused by light transmission from distant showers through very
long paths in the atmosphere.

Once the trigger channel is enabled, data stored in the memory paths are packed
into $18$-bit words, built into an event, and moved to the PC.  Note that such
a cluster readout can be enabled in two ways.  First, if some number of 
above-threshold channels are triggered in time coincidence, the stored data 
from the entire cluster will read out to the PC.  The number of channels
required for coincidence is set by an n-fold parameter in software.  Second, 
the DAQ software may make an explicit data request --- as is often useful 
during testing and baseline monitoring --- at which point all channels will 
read out for an adjustable length of time, independent of their trigger states.

As mentioned earlier, the channel baselines, thresholds, discriminator
settings, and the time coincidence number may be set dynamically from the host
PC.  In this way, the DAQ software can monitor each channel offline and
adjust the trigger parameters for the subcluster to raise or lower trigger
rates depending on drifts in background light levels.  However, the firmware
also monitors the signals online, tracking an internal baseline and variance 
for each channel by averaging the FADC outputs over $256$ clock cycles.  If the
user chooses, this fast internal baseline can be used for trigger decisions 
rather than the slower external baseline.

% -----------------------------------------------------------------------------
\subsection{PCI Readout Board}\label{subsec:pci}
% -----------------------------------------------------------------------------
When a trigger occurs in the FADC module, the data are sent to the DAQ host
computer, an Intel PC running Linux.  The host communicates with the FADC
module through a compact PCI board, a $32$~bit, $33$~MHz PCI accelerator. 
The PCI board is driven by a PCI~9056 chip from PLX \cite{Pci9056:2003ba},
and has three connections to the FADC module backplane: two high speed 
Fiber Channel cables for control and data, and a USB 2.0 link for clock.

The PLX PCI~9056 chip has several very convenient features suitable for our
application.  First, it implements a DMA engine for direct memory access
transfers into the host memory, freeing CPU resources for DAQ functions and
disk i/o.  The card actually contains two DMA channels, which we use for the
transfer of control and data.  Second, the PLX drivers allow the chip to 
operate in so-called direct slave C-Mode~\cite{Pci9056:2003ba}, in which 
locations in the PCI address space are mapped into host memory.  
This allows the DAQ host direct read and
write access to the PCI~9056 registers and addresses, greatly simplifying the
controller software and significantly reducing the overhead of DMA transfer
set ups.

% Fig 3
\begin{figure}
\centering
   \includegraphics[width=0.7\textwidth,angle=0,clip]{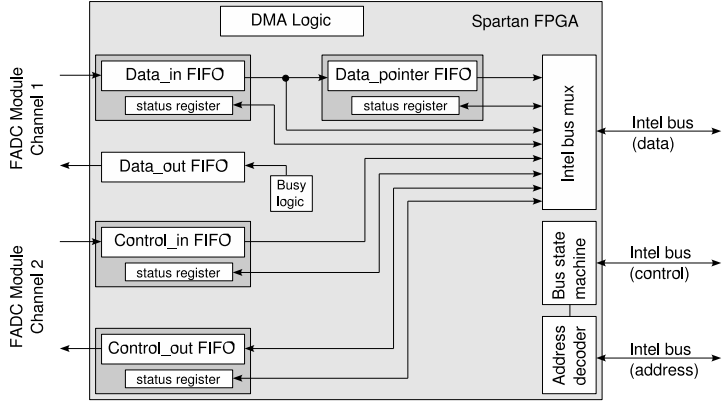}
   \caption{\footnotesize Architecture of the Xilinx Spartan
      FPGA aboard the PCI card.}
   \label{fig:spartan}
   \vspace{0.5cm}
\end{figure}

The second major component of the PCI card is a Xilinx Spartan-3 
FPGA~\cite{Xilinx:2004}, which is responsible for managing control requests
sent to the FADC module and data heading to the host PC.  To handle control and
data, the Spartan device implements four FIFOs in firmware: two for control
transfers (\texttt{control\_in} and \texttt{control\_out}) and two for data 
transfers (\texttt{data\_in} and \texttt{data\_ptr}), as shown in 
Fig.~\ref{fig:spartan}.  The control FIFOs are responsible for setting and
receiving the FADC status.  The \texttt{data\_in} FIFO contains data events sent
from the FADCs and written to a DMA buffer, while \texttt{data\_ptr} stores a 
list of buffer addresses marking the start of each event.  Finally, there is
an additional FIFO, \texttt{data\_out}, that is responsible for blocking 
transfers from the FADC module when the data FIFOs are full.

When the PLX card operates in direct slave mode, the address spaces of the
Spartan are also mapped into PC memory.  Hence, the DAQ host can operate the 
FADC module transparently and efficiently.  Module commands are sent via writes
to the control FIFO status registers; the DMA engine is initialized via direct 
writes to the PCI~9056 registers; and events are directly read from the 
data FIFO addresses.

The PLX card is nominally capable of very large data transfers, up to $8$~MB in
a single transfer at a rate of $130$~MB~s$^{-1}$.  In benchmarking tests 
conducted on a Windows PC at Nevis, we have observed sustained transfer rates 
of $\sim80$~MB~s$^{-1}$.

% -----------------------------------------------------------------------------
\subsection{Software DAQ}\label{subsec:daq}
% -----------------------------------------------------------------------------
The operation of the card and data acquisition system in the PC is fairly 
straightforward.  Its primary responsibilities are to initialize the FADC
module for data-taking; prepare the PCI card for DMA transfers; allocate
sufficient memory to store events from the FADCs; perform software-level
triggers on incoming events; and write passed events to disk.

% Fig 4
\begin{figure}
\centering
   \includegraphics[width=0.6\textwidth,angle=0,clip]{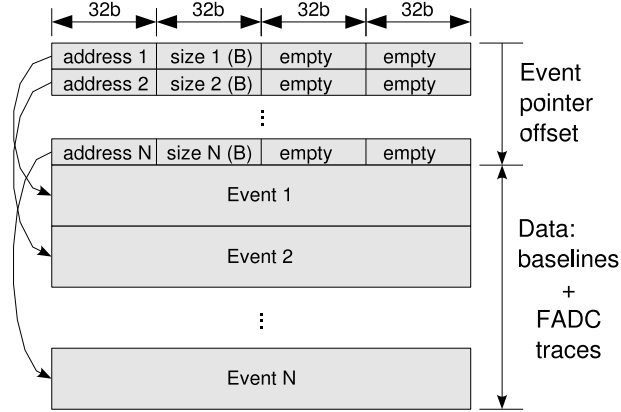}
   \caption{\footnotesize Physical layout of a DMA event buffer.}
   \label{fig:dmabuf}
   \vspace{0.5cm}
\end{figure}

\begin{figure}
\centering
   \includegraphics[width=0.5\textwidth,angle=0,clip]{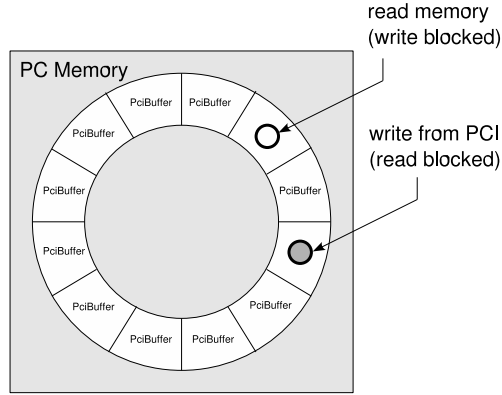}
   \caption{\footnotesize Logical setup of the software DMA ring buffer, set
      up for asynchronous readout.}
   \label{fig:ringbuf}
   \vspace{0.5cm}
\end{figure}

The DAQ host is an Intel PC running Linux kernel 2.4, chosen for compatibility
with the PLX PCI device driver.  At program startup, the DAQ software must 
allocate large, contiguous blocks of physical RAM --- in sections of up to 
$8$~MB of main memory --- for DMA transfers.  Each DMA buffer, shown in 
Fig.~\ref{fig:ringbuf}, contains a list of event pointers and actual events.
An event pointer consists of an address word and an event size word, which are
determined during processing in the FADC module and Spartan-3 FPGA.  The events
themselves contain an event type, a time base, a list of baseline averages
and variances for the entire cluster, and the actual FADC traces.  The number
of trace samples in each event, which determines the event size, is set by the
control software.

The large volume of data moving over the PCI bus into main memory requires the
host to allocate large blocks of contiguous physical memory (the PLX DMA engine
can move up to $8$~MB in one DMA transfer).  To overcome the memory allocation 
constraints set by the operating system, we patched the Linux kernel with a
video module that allows users to reserve hundreds of MB of physical RAM at
boot time~\cite{Middelink:2003}.  A small alteration to the PLX Linux driver
makes this memory available at run time to the PLX API.

% Fig 5

Since the memory allocated for DMA transfers must be safely handled by the DAQ
process, the DAQ software wraps the PLX memory allocation functions inside a 
C++ DMA buffer class.  The class constructors and destructors automatically 
allocate and deallocate blocks of reserved memory in a transparent manner, 
safely returning the memory to the operating system even after program failure.

The abstraction of memory regions allows the software to easily handle a second
important requirement of the DAQ: namely, the need to store new data as older
events are still being processed.  For this purpose, the boot-time memory 
region is divided into DMA buffer blocks, and a reference to each buffer is 
stored in a linked list, creating a ``ring buffer'' in software 
(Fig.~\ref{fig:ringbuf}).

In our implementation, we have divided the contiguous memory space of about
$100$~MB into blocks of $4$~MB to $8$~MB, matching the maximum DMA transfer
size of the PCI interface and containing about $3$ to $6$~ms of FADC data.  
These blocks are linked into a ring structure, and the DAQ operates 
by iterating through the ring, analyzing the data block-by-block.  Rather than
making synchronous read and write requests for each buffer, the software splits
reading and writing into two threads.  The first thread starts and stops DMA
writes into each buffer, while the second iterates through the ring and reads
data after they have been written.  Buffer writing is quite efficient, as the
memory mapping feature of the PCI interface allows the software to start and
stop DMA transfers simply by clearing or setting four PCI~9056 registers.
Moreover, such a data acquisition scheme automatically prevents runtime errors 
like writing into unread buffers and reading from unwritten buffers, as each 
thread blocks access to the particular buffer it is using.

When a buffer is read, the software must rapidly make a trigger decision ---
for instance, further threshold calculations, timing cuts, or geometrical
triggers on phototubes.  If the data survive the cuts, they are saved to
a RAID-1 disk and the buffer lock is removed.

% -----------------------------------------------------------------------------
\section{Discussion}
% -----------------------------------------------------------------------------
Using the partial readout system of sixteen photomultipliers, the prototype
FADC module, the PCI board, and a host PC, we have carried out several simple
light calibration tests at Nevis.  Placing the PMTs in a dark box and pulsing
them with a blue LED (attenuated with neutral density filters), we observed the
single-electron response of the subcluster.

At the typical operating gain of the phototubes ($5\times10^{4}$) we found that
one photoelectron corresponds to approximately one ADC count.  This suggests
that the dynamic range of the FADCs is sufficient to view showers in the
desired energy range without the need for additional low-gain overflow
channels.  Changes in the background light level can be accounted for by
dropping and raising the threshold and discriminator constraints in the DAQ
software.

Within the DAQ software itself, we have implemented further simple threshold
triggers to analyze the data, but we have yet to build a geometrical trigger
for a full camera (sixteen pixels is not sufficient).  To use this
readout in a full telescope will require an intermediate readout board to 
collect data from sixteen or more FADC modules.

\ack
This project is supported by the National Science Foundation under grant 
NSF-PHY-0134007.

% -----------------------------------------------------------------------------
\bibliography{manuscript}    % Produce bibliography via BibTeX.
% -----------------------------------------------------------------------------

\end{document}